\documentclass[12pt, manuscript]{aastex}

\newcommand \beq {\begin{equation}}
\newcommand \enq {\end{equation}}
\newcommand \alfven {\textrm{Alfv\'en}}
\newcommand \lint  {\lambda_{z,~int} }
\newcommand \lFWHM {\lambda_{z,~FWHM}}

\slugcomment{}

\shorttitle{Numerically Converged Amplitude of MHD Turbulence}
\shortauthors{Shi, Krolik \& Hirose}

\begin{document}


\title{What Is The Numerically Converged Amplitude of MHD Turbulence in
Stratified Shearing Boxes? }


\author{Jiming Shi\altaffilmark{} and Julian H. Krolik\altaffilmark{}}
\affil{Department of Physics and Astronomy, Johns Hopkins
University, Baltimore, MD 21218}
\and
\author{Shigenobu Hirose\altaffilmark{}}
\affil{Institute for Research on Earth Evolution, JAMSTEC,
3173-25 Showamachi, Kanazawa-ku, Yokohama, Kanagawa 263-0001, Japan}

\begin{abstract}
We study the properties of the turbulence driven by the magnetorotational
instability (MRI) in a stratified shearing box with outflow boundary
conditions and an equation of state determined by self-consistent dissipation
and radiation losses. A series of simulations with increasing resolution are
performed within a fixed computational box. 
We achieve numerical convergence with respect to radial and azimuthal
resolution.  As vertical resolution is improved, the ratio of stress to pressure
increases slowly, but the absolute levels of both the stress and the
pressure increase noticeably.  These results are in contrast with
those of previous work on unstratified shearing boxes, in which improved
resolution caused a diminution in the magnetic field strength.  We argue
that the persistence of strong magnetic field at higher resolution found
in the stratified case is due to buoyancy.  In addition, we find that
the time-averaged vertical correlation length of the magnetic field near
the disk midplane is $\simeq 3$ times larger
than found in previous unstratifed simulations, decreasing slowly with
improved vertical resolution.
We further show that the undulatory Parker instability drives the magnetic field
upwelling at several scaleheights from the midplane that is characteristic of
stratified MHD-turbulent disks.

\end{abstract}



\section{INTRODUCTION \label{sec:intr}}
The physical mechanism for transfer of angular momentum through
disk gas is believed to be the magnetorotational instability (MRI)
\citep[][]{BH91}. The MRI is a local linear instability, driven by exchange of
angular momentum along magnetic field lines
threading material at different distances from the central object. Its
growth rate is of order the orbital angular frequency, and
the fastest-growing wavelength is approximately the propagation
length of $\alfven$ waves in one orbital period \citep[][]{BH98}.
But what mechanisms limit the exponential growth of the MRI
and determine the saturation amplitude of the MRI-driven
turbulence is still a big problem to be resolved.

In the context of hydrodynamic turbulence, small scale dynamics do not
significantly affect large scale behavior. 
In magnetohydrodynamics, however, and in particular in the somewhat special
circumstances of the {\it anisotropic} turbulence expected in accretion disks
due to their characteristic orbital shear, this may not be the case.
Given the inadequacy of analytic methods, numerical simulations are required
for quantitative studies.  To reach the smallest possible length scales,
the best approach is not to study an entire disk, but only an annulus of
limited azimuthal extent.  A ``shearing box" approximation, in which
such an annulus is stretched into rectangular geometry, periodic boundary
conditions are applied along the azimuthal direction, and sheared-periodic
boundary conditions applied along the radial direction, works well \citep{Hawley95}.
Recent unstratified shearing box simulations (without a vertical component of
gravity and imposing periodic boundary conditions in the vertical direction)
have shown that the saturation level of the magnetic stress
appears to converge to zero as the grid cell width
becomes infinitesimal when there is no net vertical flux \citep[][]{FP07}.  In
these simulations, the correlation
length of the turbulent magnetic field also became smaller and smaller as the
grid scale shrank. Only if some small but non-zero dissipation is included
\citep[][]{Lesur07, FPLH07}, is convergence achieved, with the field having
some small finite value.

This result has led to considerable discussion and puzzlement.
One possible explanation is that it is due to an unphysical approximation.
Because the vertical gravity is proportional to altitude from the midplane, it
has been thought that unstratified shearing boxes are a good model for the midplane
region of the disk, where gravity should be very weak. However, eliminating
gravity also eliminates a physical length scale, the vertical pressure scale
height.
The absence of any physical length scale other than the grid scale may explain
the decay of the correlation length with diminishing grid scale, and the decay
of the magnetic field strength as well.  In this paper, we wish to test whether
the dependence of magnetic stress on grid scale changes when vertical gravity,
along with the physical length scale it introduces, are included.
We will also add a further degree of realism by explicitly
computing the temperature by balancing dissipation with radiative cooling.

We are not the first to explore MRI behavior in a stratified shearing box;
the shearing box with vertical stratification has been studied and developed
since the $1990$s. Early stratified disk study has already shown
very different vertical structures compared with the unstratified model
\citep[][]{Brand95, Stone96}.  Even though these simulations extended only
a few scaleheights from the midplane, where the structure did resemble
that of an unstratified homogeneous disk, at high altitudes, the gas density was
much lower and the magnetic field dominated the pressure.
In those simulations, a simple equation of state, either isothermal or adiabatic,
was adopted. No cooling, or only simple optically-thin thermal relaxation, was
considered. Due to the numerical difficulties of defining an outflow boundary
above an approximately hydrostatic fluid, less realistic vertical boundary
conditions were often used: for instance, stress-free and reflecting boundary
conditions were used in \citet{Brand95}, and outflow boundaries with extra
resistive layers below them in \citet{Stone96}. 
By including the displacement current in the equation of motion,
\citet{Miller00} introduced an $\alfven$ speed limiter in their simulations
in order to lengthen the time-step, providing a way to more computationally
efficient long-term simulations. Again, the equation of state was a simple
isothermal one, and no radiative transfer effects were studied. 
Using flux-limited diffusion (FLD) to solve the radiation transfer problem,
\citet[][]{Turner04} performed the first illustrative calculation of a
vertically stratified disk segment that included dissipation and radiation
effects. However, in this calculation, energy was not completely conserved:
only magnetic energy losses were captured into heat. Full energy conservation
was first achieved by \cite{Hirose06}, who simulated a gas pressure dominated
disk annulus. To smooth the field when it crosses the outflow boundaries,
they added a small amount of artificial resistivity into the ghost cells.
They also applied the FLD approximation to describe the
radiative transfer within the disk. Recently, vertically stratified disk
segments with both comparable radiation and gas pressures \citep[][]{Krolik07,
Blaes07} and radiation pressure much larger than gas pressure
\citep[][]{Hirose09} have also been studied.
In these papers,
a similar technique with thin diffusive layers extended into the problem volume
near the top and bottom boundaries was tested and implemented.
In all these stratified simulations, significant contrasts between stratified
and unstratified disks were found. For instance, magnetic buoyancy leads to a
highly magnetized `corona' \citep[e.g.,][]{Miller00, Turner04, Blaes07} that is
completely absent in unstratified disks.

As already mentioned, it is the objective of this paper to study
numerical convergence of a density-stratified shearing box
in which energy is conserved and radiation transfer is taken into account.
To do so, we performed a set of numerical simulations with increasing
resolution, but we did not include any explicit diffusivity
other than the small amount near the boundaries and the von Neumann-Richtmyer
bulk viscosity in compressive regions to
thermalize kinetic energy in shocks.

The paper is organized as follows. In $\S$\ref{sec:method}, we describe the
calculational method and initial setup of the simulations. The main results are
presented in $\S$\ref{sec:result}. In $\S$\ref{sec:conclusion} we discuss our
results and summarize our conclusions.

\section{CALCULATIONAL METHOD\label{sec:method}}

\subsection{BASIC MODEL\label{sec:model}}
We adopt the code described in \citet{Hirose06} as our basic tool
for the calculation. The complete updated (with Compton scattering
included) equation set can be found in $\S 2.1$ of \citet{Hirose09}. 
Under the ``shearing box" approximation, the disk annulus has an azimuthal 
shearing velocity $v_y=-(3/2)\Omega x$ in the background.
We carefully include the Coriolis force, gravitational tidal forces and the
vertical component of
the gravity in the momentum equation. As the
magnetic Reynolds number is usually large in accretion disks, the
ideal MHD limit is a simple but reasonable choice to describe
the magnetic field. The radiation field is described by the FLD
approximation. For simplicity, the opacity is thermally averaged. The gas
and radiation exchange both momentum and energy via Thomson scattering
and free-free absorption. Energy exchange via Compton
scattering is also included although it contributes little under the
conditions we examine here.  To complete the equation
set, we assume an adiabatic index $\gamma=5/3$ to relate the gas pressure to
internal energy. One merit of the code is that it enforces energy
conservation very well. The only violation comes from the density and energy
floors and the velocity cap.  Our density floor is $10^{-5}$ times the initial
midplane density; our energy floor and velocity cap are the same as in
\citet[][]{Hirose06}.  As discussed in
previous work \citep[e.g.][]{Hirose06,Hirose09}, the artificial energy injection
due to those three limiting values is negligible compared to the total
energy content.

\subsection{INITIAL SETUP\label{initial}}
As a test case, we chose to repeat the physical problem first investigated
by \citet{Hirose06}, a stratified shearing box in which, averaged over 50
orbits of well-developed turbulence, the gas pressure was approximately
five times greater than the radiation pressure.  Our primary reason for
using this as our test-case is that it is the only example in the published
literature of a gas-dominated stratified shearing box studied with self-consistent
thermodynamics.  When the thermal state of such a shearing box is
found self-consistently, it is fully characterized by only two parameters:
the rotation rate at the center of the box
$\Omega=5.90$ s$^{-1}$ and the surface density $\Sigma=9.89\times10^4$ g cm$^{-2}$,
producing an electron scattering optical thickness $3.3 \times 10^4$.
In terms of a Shakura-Sunyaev model, those parameters
correspond to a disk annulus at a radial distance $r=300 r_g$ ($r_g=GM/c^2$ is the
gravitational radius) from a central black hole with mass $M=6.62 M_\odot$, an
accretion rate that would yield a total
luminosity of $0.1$ of the Eddington rate if the efficiency is $10\%$ in
rest-mass units, and a Shakura-Sunyaev stress ratio $\alpha=0.03$. The predicted
effective temperature is $T=4.77\times 10^5$ K. In the gas pressure-dominated
limit, the scale height (half-thickness) of the disk is $H=3.53\times 10^6$~cm.
We choose the characteristic scale height $H$ as our length unit.  

Our inital condition is similar to the one used in
\citet{Hirose06} except that we assumed a dissipation profile
$dF/dz\propto d\Sigma^{1/2}/dz$ instead of $\propto d \ln\Sigma/dz$.
This form for the dissipation profile is more similar to the
time-averaged dissipation profile found in previous simulations,
so it helps the simulation to pass the transient phase quickly,
but does not affect the later stages.
The initial disk is in approximate
hydrostatic and radiative equilibrium below the photosphere.
Outside it, the flux is constant, and the gas density is set to the
density floor value. The
initial configuration of the magnetic field is a twisted azimuthal
flux tube of cicular cross section, centered
at $x=z=0$ and having a radius of $0.75H$. The tube has uniform interior
field strength $B_0=2.36\times 10^6$~G, corresponding to $4\%$ of
the initial box-averaged gas pressure plus radiation pressure. The
maximum poloidal component of the field strengh is $5.90\times
10^5$~G, while the net vertical flux is zero.
Although this initial magnetic field has both net azimuthal flux
and net helicity, neither quantity is preserved, and therefore neither
provides any long-term constraint on the evolution of the magnetic field.
The fastest-growing vertical
MRI wavelength in the midplane in the initial state is $1.90\times 10^6$~cm,
which is resolved with $8.6$ grid cells in our lowest resolution case.
We use this initial condition for all simulations in this work. The
calculation is begun with a small random perturbution in the
poloidal velocity. The maximum amplitude of each velocity
component is $1\%$ of the local sound speed defined as
$c_s\equiv\sqrt{(4E/9 +\gamma p)/\rho}$, where $E$ is the radiation
energy, $p$ is the thermal pressure and $\rho$ is the gas density.

All simulations in this paper start with the same
computational box, which extends $2H$ in the radial direction, $8H$
along the orbit, and $8H$ on either side of the midplane.
We first performed a standard run, denoted as STD32,
with moderate resolution: $32\times64\times256$ cells ($x \times y \times z$)
with constant cell size $\Delta x=\Delta z= H/16=\Delta y/2$.
Next we carried out three runs that had doubled resolution in only one of the
three directions; they are labelled: X64 for
$64\times64\times256$, Y128 for $32\times128\times256$ and Z512
for $32\times64\times512$.  Finally, we raised the resolution by
doubling the number of cells in all three directions, and the
corresponding run is called DBLE. It is the best resolution
we can reach in practice; further simulations with smaller cells
will require both a higher efficiency code and more powerful computational
facilities. The details of the runs we performed are listed in Table \ref{tab:1}.

The azimuthal boundaries are purely periodic while the radial
boundaries are shearing-periodic \citep[][]{Hawley95}. At the top
and the bottom surfaces, the boundary conditions are outflow
(free) boundaries as used in \citet[][]{Hirose06}. Similarly, we
introduce a small resistivity in cells adjacent to the top and
bottom of the box in order to prevent magnetic field
discontinuity. This resistivity is tapered sinusoidally from its
maximum value in the ghost zone, $0.005\times min(\Delta
x^2,\Delta y^2, \Delta z^2)/\Delta t$, to zero at $1H$ into the
computational domain, i.e. $16$ cells into the problem volume for
STD32, X64 and Y128, and $32$ cells for Z512 and DBLE. Here $\Delta t$ is the
time step in simulations.

The box height is designed to be large enough to contain most of the gas for
long-term evolutions. However, the surface mass density of the disk segment can
vary due to matter added by the density floor or, more importantly
here, gas outflow. In this paper, we permitted variations of the
surface mass density only up to $2.5\%$. In Z512 and DBLE, we enlarged the box
height by a factor of $1.25$ to prevent outflow when the matter
swells\footnote{Once the loss of surface density becomes greater than $2.5\%$, we go
back several orbits and perform the resizing on the restart file of that time. The
simulation is then continued with the enlarged box.}.
When doing so, we kept the resized box at the same resolution
as before, i.e., $\Delta x$,$\Delta y$ and $\Delta z$ were unchanged.
When the new cells were initialized, they were given density
and energy equal to the floor values. The velocities were set to zero except for
the background shear.
The transverse components of the magnetic field were copied from the values
in the vertically-aligned cells on the old boundaries , while
the perpendicular component was calculated by enforcing the divergence-free
constraint.  The layers with artificial resistivity were resized by the same
factor of 1.25. We configured the resizing this way to keep the surface mass
density of the box nearly constant while not introducing large pressure
gradients or energy injection in the extended zones (the added energy was $<
1\%$ of the total).
We also tested whether this size change alone causes any noticeable effects
on the results.  A test run was started at $t=90$ orbits of STD32 with its height
enlarged by a factor of $1.25$ and evolved $50$ more orbits. It was then compared
with the data from $t=90$ to $140$ orbits in STD32. They are statistically
similar: the time- and volume-averaged stresses and energies after resizing
differ by $\leq 3\%$ from the standard.  Note that there is no significant gas
outflow in STD32 during this test period, so the box resizing must be
responsible for any changes. In all runs, we chose a duration long enough that
the staturation level appears to be quasi-steady. Depending on the simulation,
this required between $\sim 120$ and 300 orbits, which amount to $\sim 20$
--$40$ cooling times.

\section{RESULTS \label{sec:result}}

We are primarily interested in time-averaged quantities representative
of conditions in a steady-state disk.  To determine the appropriate
time-averaging period, we examined how rapidly the effects of
our initial conditions and transient response die away.  As Figures~\ref{fig:STD},
\ref{fig:others}, and \ref{fig:alpha}
show, in every simulation the various contributions to the stress reach
levels characteristic of the statistical steady-state by $\sim 10$ orbits
from the beginning.  Topological properties of the initial magnetic
field are erased rapidly: the net azimuthal flux changes sign on $\sim 5$~orbit
timescales, and the volume-integrated magnetic helicity has a correlation
time of at most $\sim 3$ orbits.  We therefore defined each simulation's
time-average as beginning at 10 orbits and running to its end.

\subsection{STANDARD RUN\label{sec:result_1}}
\begin{figure}
\plotone{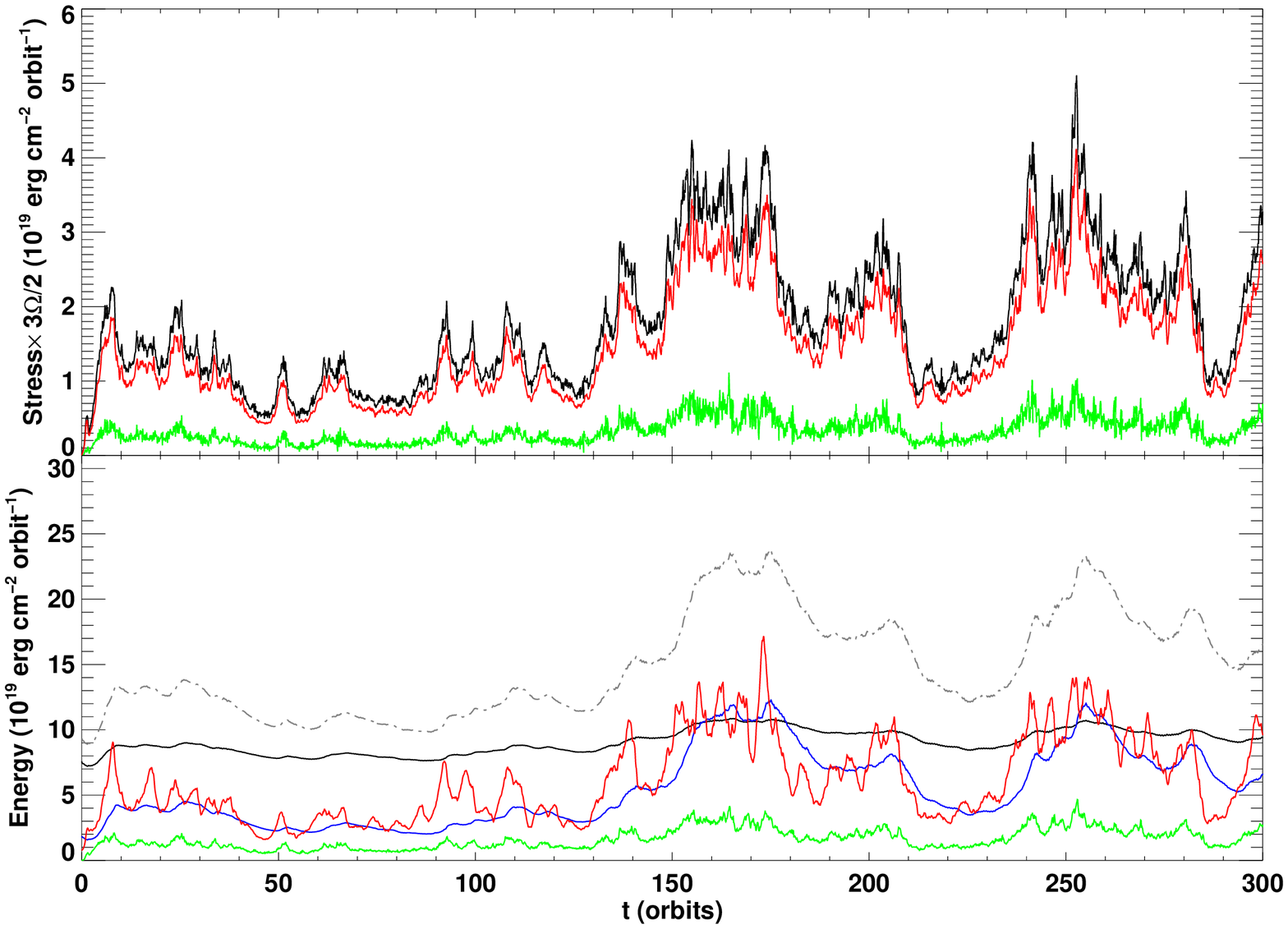} \caption{\small{Volume-integrated stresses (upper panel) and
energies (lower panel) in STD32 as a function of time. \textsl{Top panel}:
Maxwell stress (red), Reynolds stress (green) and the sum of these two (black).
\textsl{Bottom panel}: internal energy (black), radiation energy (blue), $10$
times the magnetic energy (red), $10$ times the kinetic energy(green) and
total energy (grey dash-dotted). The quantities are all
vertically integrated and the stress is multipled by $3\Omega/2$.}\label{fig:STD}}
\end{figure}

We show the time evolution of the stresses and energies of STD32
in Fig.\ref{fig:STD}. STD32 has the same resolution and nearly
identical initial setup as the run in \citet{Hirose06}, which makes
it easy to compare them statistically (note, however, that STD32 ran
for five times as long). STD32 appears to
have two stages after the transient decay, which separate around $130$
orbits. During the first stage, the maximum Maxwell stress is
about $3$ times as great as the minimum, which is nearly the same
as the gas pressure-dominated run in \citet{Hirose06}. The ratio of
box-integrated stress to box-integrated total pressure (including
both the gas and radiation pressures) is $0.015$, and that is also 
consistent with the value found in \citet{Hirose06} (it was $0.016$ in
their paper).  In the second stage, the peak value of the stress
doubles, and the range of the fluctuations
is roughly a factor of $4$. A quick look at the radiation
energy and gas energy plots in Fig.\ref{fig:STD} reveals that although the
disk starts as a gas pressure-dominated system, it gradually
evolves toward a situation with a larger ratio of radiation
to gas pressure: the time- and volume-averaged ratio
$\langle\langle p_{rad}\rangle/\langle p \rangle\rangle$ is $\sim
0.2$ for the first $130$ orbits, but increases to $\sim 0.4$ for
the rest with a variation range $\simeq0.2 - 0.5$.  Here the first
(inner) $\langle\rangle$ represents a volume average and the
second (outer) denotes a time average. Comparison with the previous
simulation with comparable gas and radiation pressure
\citep{Krolik07} is helpful. In that run, the variation of the
stress is $\sim 6$, which is slightly bigger than that of
the second stage of STD32; the nominal time-averaged
$\alpha$-parameter is $\simeq 0.03$, while it is a bit smaller, $\sim
0.02$, for STD32; the ratio of radiation pressure to gas pressure
varies over the range $\simeq 0.5 - 2$, which is beyond that of
STD32.
Despite large fluctuations, STD32 clearly achieves a
quasi-stable stage for the last $150$ orbits. All
other four runs in this paper also show the feature of increasing
$\langle p_{rad}\rangle/\langle p \rangle$, and they
are terminated when a quasi-stable stage like the one in
STD32 is reached.

\subsection{CONVERGENCE \label{sec:result_2}}
\begin{figure}
\plotone{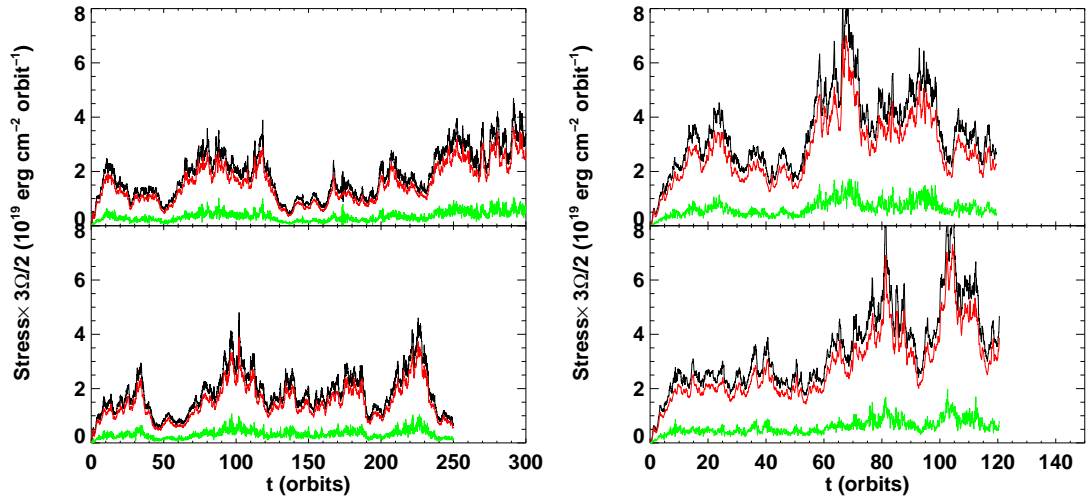} \caption{\small{Volume-integrated stresses in Y128(upper left
panel), X64(lower left panel), Z512(upper right panel) and DBLE(lower right
panel) as a function of time. The color curves are coded as in the top panel of
Fig.\ref{fig:STD}. }\label{fig:others}}
\end{figure}

\begin{figure}
\plotone{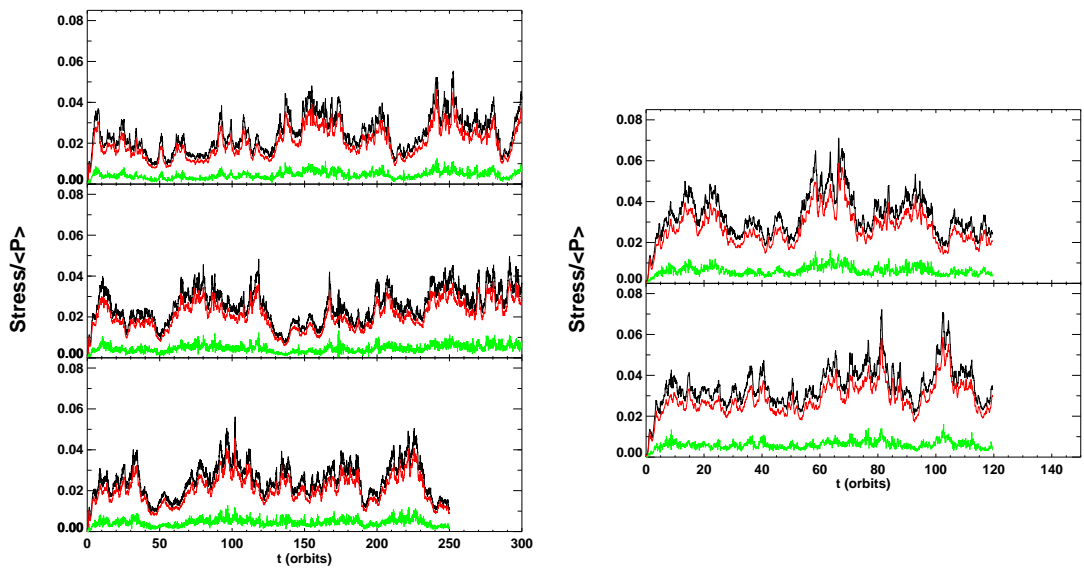} \caption{\small{Volume-integrated stresses in ratio to
simultaneous volume-averaged total pressure for STD32, Y128, X64 (left, from
top to bottom), Z512 and DBLE (right, from top to bottom).
Color code is the same as in Fig.\ref{fig:others}.}\label{fig:alpha}}
\end{figure}

The saturation states of the other four runs are illustrated in
Fig.\ref{fig:others}. Clearly the saturation level of X64 and Y128 (left two
panels in the graph) are similar while the Z512 and DBLE (right two panels) have
relatively higher mean values. The offset between the left two and the right two
begins right after the transient decay. It grows even larger after the first
$60$ orbits in both Z512 and DBLE: at that point, the radiation energy becomes
comparable with the gas energy.

To study convergence quantitatively, we need not only temporal and spatial averages
of the stress, but proper ways to normalize it.  Here we choose the averaging
time to be from the end of the transient phase ($10$ orbits) to the end of the run; the
volume-integral of the stress is used as the spatial
average. Unlike the nonstratified case, the pressures in our simulations show
both consistent spatial gradients, depending on height from the midplane, and
significant trends over time.  We therefore employ three
different methods of normalization, and examine the quality of numerical
convergence in each case. 

We first normalize the different kinds of stress and energy by the initial
volume-averaged total pressure $P_0$ (sum of gas pressure and radiation
pressure): $P_0=9.36\times10^{11}$ ergs cm$^{-3}$ in our simulations. This
normalization definition is extensively used in unstratified shearing box simulations,
but with the total pressure replaced by thermal pressure alone.
The time- and box-averaged values are given in Table \ref{tab:1}.  Scanning
across each line, one can see which quantities are sensitive to resolution;
in general, convergence has clearly been reached with regard to $x$ and $y$
cell size, but not with respect to $\Delta z$.
The normalized Maxwell stress is constant at $\simeq 0.03$ when the resolution in the
radial or azimuthal directions increases, but its value is almost doubled when the
vertical resolution is raised by a factor of $2$.  Similarly, the
magnetic energy and turbulent kinetic energy rise by about a factor of two when
the vertical cell count is doubled, but are independent of the horizontal cell dimensions.
By contrast, in unstratified simulations, when the resolution improves, the
saturation level either decreases toward zero with a zero net-field configuration
\citep[][]{FP07,Simon09} or increases weakly for mean azimuthal field models
\citep[][]{Guan09}. We have net azimuthal field in our simulations, but it is
not fixed, and even the sign of the net azimuthal flux changes.

A second useful normalization standard is the horizontal average
of the time-dependent total pressure in the midplane.
The time-averaged values for the stresses and energies normalized in this fashion are
listed in Table \ref{tab:1}, too.  They depend on resolution in a way very similar
to the ones using the absolute normalization except that their values are almost
one order of magnitude smaller.
Again, in this normalization there is no dependence on horizontal resolution,
but increasing resolution in $z$ leads to larger values. For example, the
Maxwell stress for DBLE is almost twice that of STD32.

Considering that $P_0$ is just an arbitrary initial guess for the total pressure,
and $P(0)$ does not reflect the properties of the whole box, it is more physical
to normalize the energies and stresses to the simultaneous volume-averaged
total pressure, i.e., $\langle P\rangle\equiv \langle p_{rad}+p\rangle$. The
time evolution curves of the stress ratios using this normalization
are plotted in Fig.\ref{fig:alpha}.
Compared to the absolute stresses shown in Fig.\ref{fig:STD} and \ref{fig:others}, the
stress ratios after normalization show considerably smaller peak-to-peak
variations.  Normalized in this way, there is also a significantly smaller increase
when the resolution along the vertical
direction increases. The time averaged values for this normalization are the
third group in Table~\ref{tab:1}. The normalized Maxwell stress is 
$\sim 0.02$ for STD32, X64 and Y128, and increases only to $\simeq 0.03$ for Z512
and DBLE.  This sort of normalization is the best of the three to use for
estimating the Shakura-Sunyaev $\alpha$ parameter because it makes use of the
actual volume-integrated total pressure, and we see that with the best resolution
employed here, we are approaching convergence in defining its value.  We emphasize,
however, that one can speak of a single value for this number only in terms of
a particular location in the disk and after both a
vertical integration and a time average that encompasses many thermal times.

To summarize this section, we find that numerical convergence with respect to
resolution in the $x$ and $y$, but not $z$, directions has been achieved
for the absolute values of stress and energy in a stratified shearing box.
Increasing $z$ resolution at the level we have reached leads to rising absolute
values of stress and pressure.  On the other hand, we come close to reaching
convergence with respect to all three sorts of
resolution for the ratio of stresses and energies to the time-dependent
total pressure.  In the next subsection, we show that certain detailed
features of the magnetic field show similar convergence properties and
cast light on why stratified shearing boxes differ from unstratified.

\subsection{FIELD STRUCTURE \label{sec:result_3}}

\begin{figure}
   \plottwo{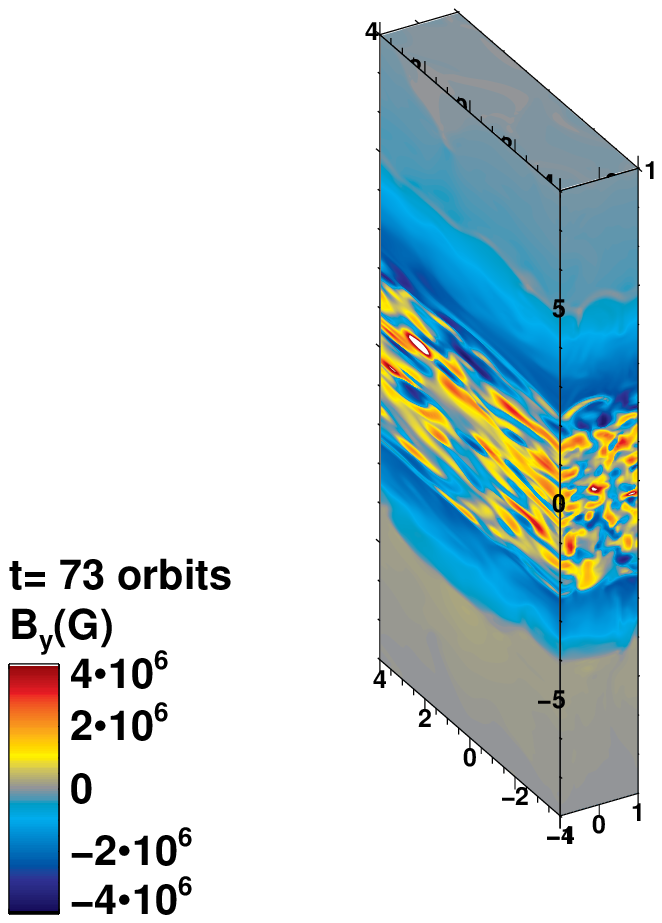}{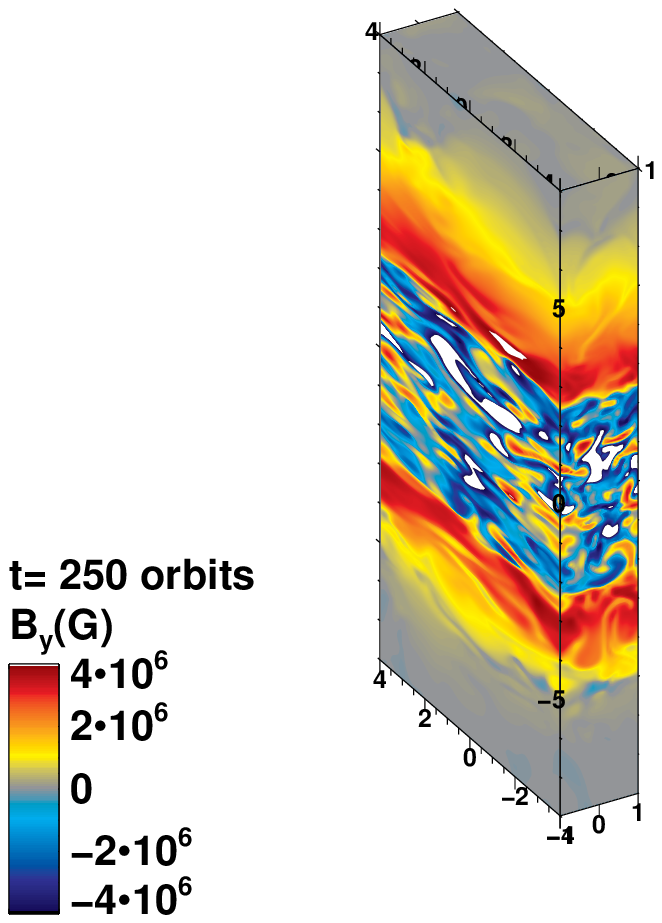}
\caption{\small{Snapshots of the $B_y$ distribution on the box surfaces at
$t=73$ orbits and $t=250$ orbits in STD32.}
\label{fig:by}}\end{figure}

In order to investigate the effect of stratification on the magnetic field, we consider
the vertical correlation length of the field. As the azimuthal component dominates
in our simulations, we restrict our attention to $B_y$.  We
present two 3-D snapshots of the distribution of azimuthal field strength in STD32
in Fig. \ref{fig:by}. In the plot, color contours of the field strength are
mapped onto the surfaces of the box.  For contrast, we choose a pair of times, one
($t=73$ orbits) when the magnetic energy is low and one ($t=250$ orbits) when
it is high.  We find the field distribution below $\pm 2 H$ of height is distinct
from that above: within that distance of the midplane, the field is turbulent,
whereas at higher altitudes it is much more regular.  Far from the midplane,
long filamentary regions of relatively smooth $B_y$ with the same sign extend more than
1--$2 H$ beyond the disk core.  The field becomes disordered again above
$\sim \pm 5 H$, where the sign of $B_y$ sometimes flips.  The
same features are consistently observed in all our other runs.  Stratification
clearly has a strong effect on the qualitative organization of the magnetic field.

This fact leads us to ask if stratification also influences the field
structure of the midplane region. 
Let us first calculate the physical scale height by taking $\langle c_s/\Omega\rangle$.  
The time averaged scale height is then denoted as $\mathcal{H}$ hereafter.
This scale height is a better measure
of the physical scale length of the disk than our guessed unit of length, $H$.
The physical scale height varies within
the range $\simeq 1.4$--$2H$ for STD32, with time average $1.65 H$ (similar variations
and average values are also found in X64 and Y128, see Table \ref{tab:1});
its range of variation is slightly wider,
$\simeq 1.4$--$2.3H$ in Z512 and DBLE, and the average is also a bit larger,
$\sim 2.0 H$.

We now calculate the vertical
correlation length of $B_y$ in the disk core, which we define as $|z| \leq H$, i.e.
within roughly half a physical scale height of the midplane. This choice allows us to
compare our results with those of \citet[][]{FP07}, whose box
extended one physical scale height vertically. Note that our DBLE run has
exactly the same resolution as their STD64 (both are $\mathcal{H}/64$
vertically). The two-point correlation function of $B_y$ at $t=t_0$
is defined as:
\beq C_z(B_y; t_0,l_z)\equiv \frac{1}{L_xL_y}\int\!\!\!\int dx dy~\frac{\int
B_y(t_0,x,y,z)B_y(t_0,x,y,z-l_z)dz}{\int
B_y^2(t_0,x,y,z)dz},\label{eq:cfunc} \enq
where $C_z(B_y;t_0,l_z)$ denotes the vertical correlation of $B_y$ at separation
$l_z$ as a function of time, and $L_x$ and $L_y$ are the box sizes along the
$x$ and $y$ directions. Following \citet{FP07}, we define an integrated
correlation length as the integral over different separations:
\beq \lint(B_y;t_0)=\int C_z(B_y;t_0,l_z)~dl_z.
\label{eq:corrlen}
\enq
Note that in \citet{FP07} the correlation length is calculated only in the
$x-z$ plane.  We find that our value of the integrated correlation length
would change by only $\sim 3\%$ (see Table \ref{tab:1}) if we had used their
definition instead of ours.  We also calculate the correlation
length defined as the full width at half maximum (FWHM) of $C_z$, i.e.,
\beq
\lFWHM(B_y;t_0)=\textrm{FWHM~of~} C_z(B_y;t_0,l_z).
\enq

\begin{figure}
\plotone{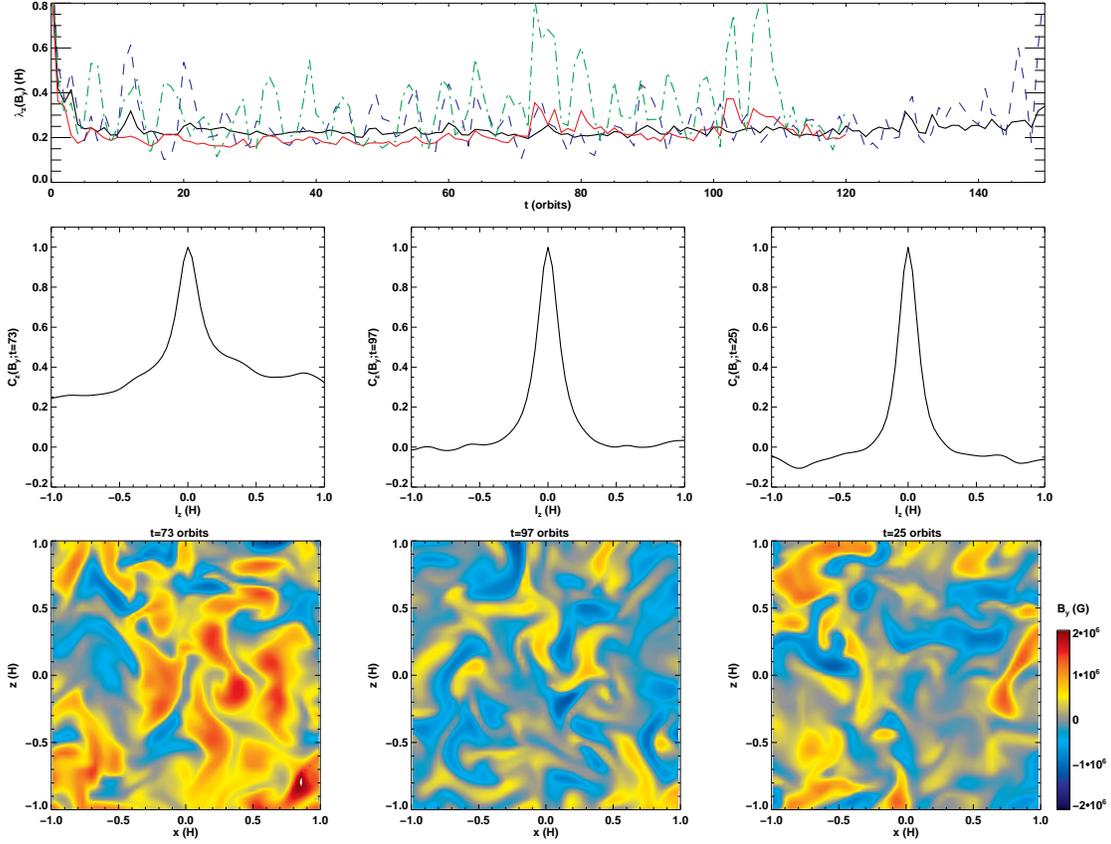} 
\caption{\small{\textsl{Top}: Radially averaged vertical
correlation length of $B_y$ for STD32 (black solid for $\lFWHM$ and
blue dashed for $\lint$) and DBLE (red solid for $\lFWHM$ and green
dash-dotted for $\lint$) as a function of time; \textsl{Middle}: The
correlation function at $t=73$, $97$ and $25$ orbits of DBLE; 
\textsl{Bottom}: $B_y$ contours from DBLE in $y=0$ plane at $t=73$, $97$ and
$25$ orbits.} 
\label{fig:corr}}
\end{figure}

We show the time histories for several definitions of $\lambda_z$ in STD32 and DBLE
in the top panel of Fig. \ref{fig:corr}.  As this figure shows, both the integrated
and the FWHM correlation
length vary with time, but $\lint$ has considerably larger fluctuations than
$\lFWHM$. A close look at the correlation function explains the discrepancy. In
the middle of the same graph, three instantaneous correlation functions, at $t=73$,
$97$ and $25$ orbits in DBLE are presented. At $t=73$ orbits, $C_z(B_y)$
clearly shows positive correlation even at large separations, making the
integrated correlation length $\lint$ as large as $0.8 H$,
about twice the $\lFWHM$ found at the same time.  On the other hand, at
$t=25$ orbits, the correlation function turns negative when $|l_z|>0.3 H$, and
thus a dip of $\lint$ ($\sim 0.1 H$) is observed at $t=25$
orbits, when $\lFWHM$ is $\sim
60\%$ large.  The two measures of $\lambda_z$ become comparable to
each other when the wings of $C_z(B_y)$ asymptote to zero at larger
separations.  For example, at $t=97$ orbits, when this occurs,
$\lFWHM$ is similar to, but slightly less than $\lint$.  We can test whether
$\lint$ or $\lFWHM$ is the more realistic measure of the correlation length by
looking at snapshots of $B_y$, such as those shown under each correlation plot.
There is no apparent change in the lengthscales of magnetic field features
between $t=25$ and $97$ orbits, despite the factor of two change in the
intergrated correlation length between those times.  The FWHM correlation
lengths at these two times are nearly the same.  Thus, we claim that
the FWHM definition is a better measure of the correlation length than
the integrated version: it corresponds more closely to one's visual impression
and varies less in time.  The large fluctuations in $\lint$ appear to be
due to sensitivity to the tails of $C_z$.

As shown by the time averaged values listed in Table~\ref{tab:1},
$\lambda_{z,~FWHM}$ decreases slightly in magnitude when the
resolution is doubled.  We find $\langle\lFWHM\rangle_t\sim 0.26 H$ for
STD32, but drops to $\sim 0.22 H$ once the $z-$cell number is doubled.
This change corresponds to a $30\%$ decrease if we scaled the length with
the physical scale
height $\mathcal{H}$: $\langle\lFWHM\rangle_t\sim 0.16 \mathcal{H}$ for STD32 and
$\sim 0.11 \mathcal{H}$ for DBLE. As we mentioned before, the
DBLE run possesses the same resolution as STD64 in \citet{FP07}, in which the time
averaged correlation length using their definition (an
integrated one) is $0.06 \mathcal{H}$; with the same definition,
the correlation length of DBLE is $\sim 0.18\mathcal{H}$, approximately triple theirs.

In sum, we find comparatively large scale ($\gtrsim 1H$) structures in $B_y$ more
than $\simeq 2H$ from the midplane, and even in the region quite close to the
midplane, features several times larger than found in unstratified simulations.
We also find that the FWHM definition of the vertical correlation length
may be more useful than the integrated one.  In terms of its degree of numerical
convergence, $\lambda_z$ is similar to the ratio of stress to simultaneous pressure:
at the resolution scales achieved in these simulations, it appears to be close
to convergence, but not quite there, particularly with respect to resolution in
the vertical direction.

\subsection{MAGNETIC BUOYANCY\label{sec:buoyancy}}
\begin{figure}
\plotone{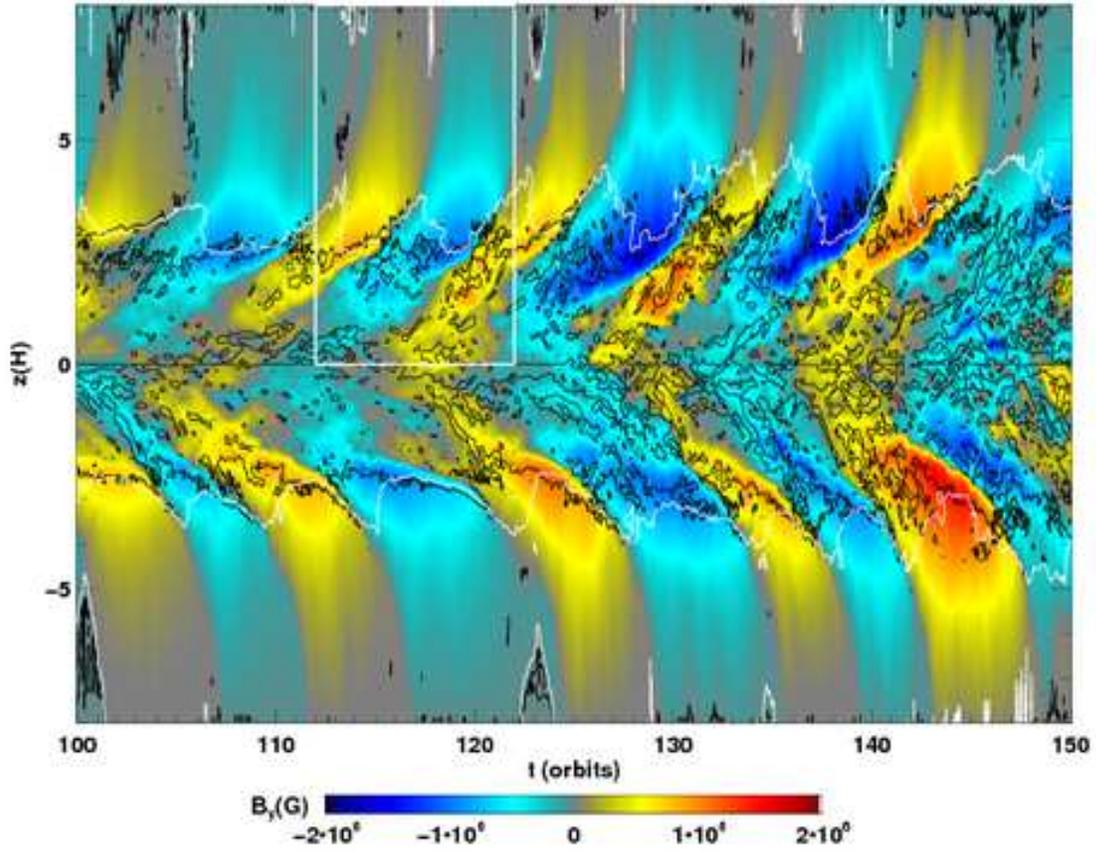} \caption{\small{Spacetime diagram of the azimuthal component
of the magnetic field. The contour lines on top delineate where and when
the magnetic Brunt-V\"ais\"al\"a frequency squared (black) changes from positive (inside
the contours) to negative (outside the contours) and the plasma $\beta$ (white) increases
from less than unity (outside the contours) to greater (inside the contours). The
white rectangle shows the section magnified in Fig.\ref{fig:bspeed}.}\label{fig:buoyancy}}
\end{figure}

As discussed in the last subsection, large filament-like structures of magnetic field
emerge above the core of the disk and extend vertically for another $1-2 H$.
These features have also been observed in many previous simulations
\citep[e.g.,][]{Turner04, Hirose06}.  In Figure~\ref{fig:buoyancy}, we
show a space-time diagram of the horizontally-averaged azimuthal component
of the field from a 50-orbit segment of
STD32\footnote{As the default dumping rate for STD32 is only one time per orbit,
to obtain this data we carried out a special high dump rate run by restarting
STD32 from $t=100$ orbits, running for 50 orbits, and dumping physical
quantities every 0.1 orbits. These data therefore do not follow STD32 exactly,
but are physically and statistically equivalent. In that sense, we still call it
STD32 here.}.
There are roughly ten episodes of field upwelling during this sample, corresponding
to a period of $\sim 5$ orbits for the process.  The sign of $B_y$ alternates
in successive events.  The upward pattern speed is $\sim 0.5 H $
per orbit at the base of the plume ($\sim\pm 2H$), accelerating to $\sim 1-2H$
per orbit near the top of the disk ($\gtrsim\pm 4H$).
These events are almost symmetrical about the midplane except for some small phase
offsets and intensity variations. 

Computing the simple
hydrodynamic Brunt-V\"ais\"al\"a frequency for our data would suggest that
the disk is always stable against buoyancy-driven instability below $\sim 2H$,
and remains mostly stable at higher altitudes, including the upwelling regions.
However, including magnetic forces leads to a very different result.

There are two magnetic buoyancy modes to consider: the interchange mode
and the undulatory Parker instability.  The former mode
is not present in our simulations: evaluating its dispersion relation with
horizontally averaged data shows that the disk is stable at nearly all times and
in nearly all locations.  To demonstrate that the undulatory mode does seem
to be responsible for these repeated episodes of magnetic upwelling, we
begin by writing down expressions for the square of the generalized magnetic
Brunt-V\"ais\"al\"a frequency for this mode in two limits: fast and slow
radiation diffusion \citep[][in preparation]{Tao09}:
\beq
N^2_{mag, fast}=\frac{g v_A^2}{c_i^2+v_A^2}\frac{d\ln{|\textbf{B}|}}{dz},
\label{eq:n2_fast}
\enq
\beq
N^2_{mag, slow}=N^2_{hyd} + \frac{g v_A^2}{c_t^2}\frac{d\ln{|\textbf{B}|}}{dz}.
\label{eq:n2_slow}
\enq
Here $N_{hyd}$ is the hydrodynamic Brunt-V\"ais\"al\"a frequency for a thermally
coupled gas and radiation mixture \citep{Blaes07}, $g=z\Omega^2$ is the vertical
component of gravity, $v_A$ is the $\alfven$ speed, and $|\textbf{B}|$ is the
magnitude of the field strength. The isothermal sound speed in the gas is
$c_i=(p/\rho)^{1/2}$, and the total adiabatic sound speed is 
$c_t=[\Gamma_1(p+E/3)/\rho]^{1/2}$, where $p$ and $E$ again are gas pressure and
radiation energy density, and $\Gamma_1$ is Chandrasekhar's generalized adiabatic
constant \citep[][]{Chandra67}.  The ``slow diffusion" limit describes the case
in which the growth rate of the
instability exceeds the photon diffusion rate so that photons are dynamically
well-coupled to the fluid; in the ``fast diffusion" limit, photons diffuse
rapidly compared to the instability growth rate, so that there is no radiation pressure
response to the mode and the perturbations in the gas are isothermal.
Note that hydrostatic equilibrium with no
magnetic tension forces is assumed in both expressions for $N^2_{mag}$.  Evaluating
the growth rate using the wavelength of the fastest-growing mode in the rapid
diffusion limit (equation A15
of \citet[][]{Blaes07}), we find that, for regions above $\sim \pm 3.5H$, 
the fast diffusion limit is the appropriate one, while at lower altitudes
the slow diffusion limit generally applies.  In approximate terms,
$N^2_{mag}$ can then be treated as the proper combination of the frequency squared
under those two limits.  The fastest-growing wavelength is always well
confined in the box and well resolved numerically in most regions except
near the midplane.

\begin{figure}
\epsscale{0.8}
\plotone{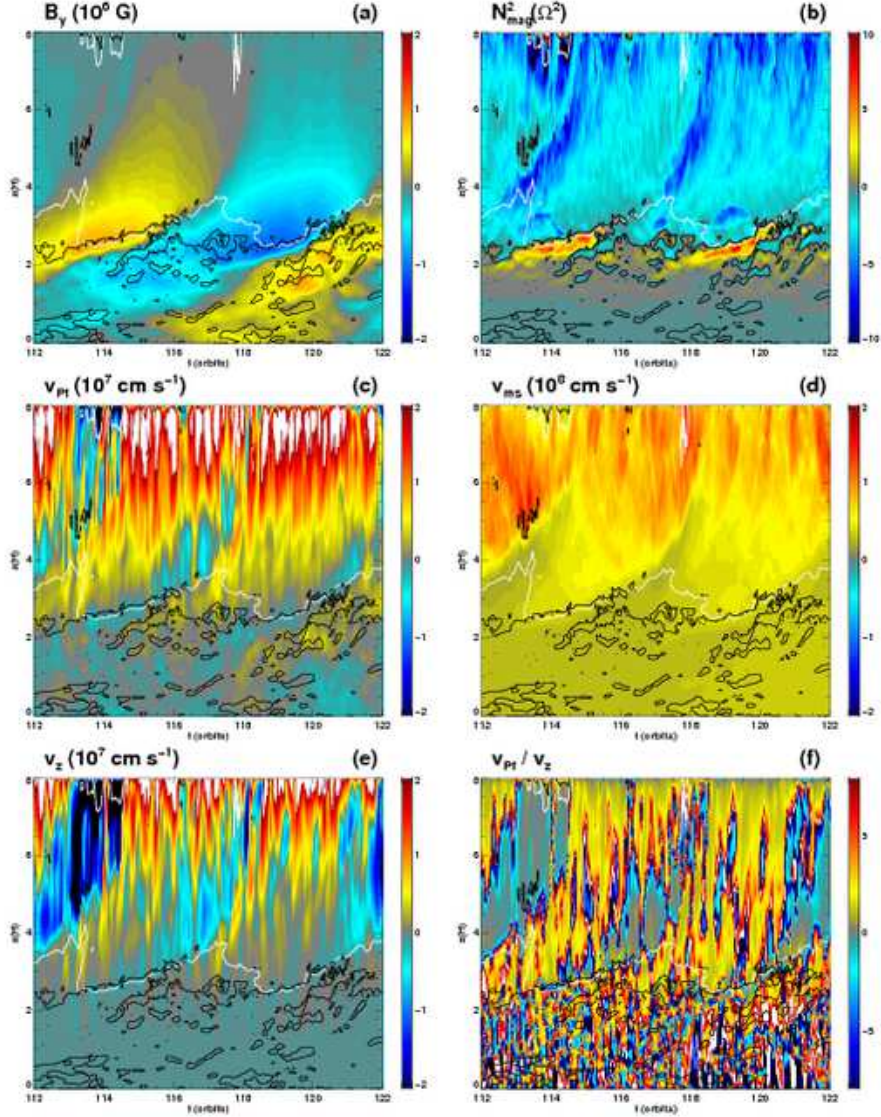} \caption{\small{Magnified view of a portion of the spacetime
diagram for STD32: (a) azimuthal magnetic field strength; (b) magnetic
Brunt-V\"ais\"al\"a frequency squared; (c) vertical velocity of the Poynting
flux; (d) magnetosonic speed; (e) vertical gas velocity; (f) ratio of the
Poynting flux velocity to the vertical gas velocity. See legend on the top of
each panel for units. The black and white contour lines are the same as in
Fig.\ref{fig:buoyancy}.}\label{fig:bspeed}}
\end{figure}

We plot the zero-frequency contour of the combined $N^2_{mag}$ (black curves)
in the space-time diagram of Figure~\ref{fig:buoyancy}. Instability takes
place only outside the contour lines.  Inside $\pm 2H$, i.e. the core
disk region, this mode is generally stable, although sometimes only marginally so.
Small unstable patches exist in the core region, and most of them are elongated in a
way suggesting buoyancy, but they exist only briefly.  The magnitude of the growth
rate in this region is small (see panel b of Fig.~\ref{fig:bspeed}), and the
implied rise speeds
are relatively slow.  These small episodes of unstable buoyancy may help explain 
why we find larger scale features in our stratified simulations than have been found
in unstratified simulations, but the fact that the wavelength of the fastest-growing
modes exceeds the size of the box suggests that this explanation probably cannot
completely answer the question of the origin of our larger correlation length.
On the other hand, $N^2_{mag}$ is negative almost everywhere at altitudes above
$\sim \pm (2$--$3) H$.  At these higher altitudes, the undulatory Parker mode
is almost always unstable with a linear growth rate $\sim (1$--$2)\Omega$.

In order to show the operation of this mechanism in greater detail, we study
a magnified view of a small domain of the spacetime diagram.  Data from the
region within the white rectangle in Figure~\ref{fig:buoyancy} are displayed on
a larger scale in Figure~\ref{fig:bspeed}. As discussed, the rising pattern is
roughly symmetric about the midplane, so by choosing a sample from the upper
half of the disk we should not lose any interesting physics. The time range of
this sample is $10$ orbits and contains almost two complete buoyancy episodes.
In this figure, we plot not only $B_y$ and $N^2_{mag}$, but also three
characteristic speeds:
the gas vertical velocity $v_z$, the magnetosonic speed $v_{ms}$,
and the Poynting flux speed $v_{P\!f}$. The last one is defined as
$v_{P\!f}\equiv 4\pi S_z/B^2$, where $S_z$ represents the $z$ component of the
Poynting flux. Thus, positive $v_{P\!f}$ means field energy is transported
upward and corresponds to field ascending. The Poynting flux speed can also be
written as $v_{P\!f}=v_z-(\textbf{v}\cdot\textbf{B})B_z/|\textbf{B}^2|$; this
form shows how, even in the MHD limit, the gas and magnetic velocities can
differ to the extent that the fluid can slide along field lines.

Panels a, c, and e of Figure~\ref{fig:bspeed} present an enlightening contrast:
although the magnetic field spacetime diagram consistently shows features in
$B_y$ rising upward from near the midplane all the way to the top of the box,
the $v_{P\!f}$ and $v_z$ illustrations demonstrate that consistent upward field
and fluid motion begins only at $\simeq 3H$ above the midplane.  In fact, the
bottom edge of the true upwelling region corresponds closely to the lower boundary
of the Parker instability region.  Thus we see that the apparent motion of
regions of strong field in the disk core and lower corona is a wave or
pattern motion almost completely divorced from genuine mass or field transport.
In fact, the pattern speeds observed in the simulation are close to the
Alfv\'en speed, suggesting that the waves in question are either Alfv\'en
or slow magnetosonic waves.  The latter mode appears to be particularly
important for short-wavelength, high-frequency variability, for which
a distinct anti-correlation between magnetic and gas pressure fluctuations
can be seen.

As already remarked, genuine rising motion of magnetic field and gas is closely
associated with the onset of magnetic buoyancy instability.  Further evidence
that undulatory Parker mode instability is the origin of upwelling is provided
by the fact that the regions of greatest growth rate (the dark
blue strips in the $N^2_{mag}$ panel) coincide with the leading edges of regions
where both $v_{P\!f}$ and $v_z$ are large.  In other words, the acceleration
to high upward velocity takes place where the instability grows most strongly.

Although true upward motion certainly takes place at altitudes several scale heights
from the midplane, it is in an absolute sense relatively slow.  Typical speeds for both
$v_{P\!f}$ and $v_z$ are $\sim (0.08$--$0.8)\mathcal{H}\Omega$.
Both are subsonic, only $\sim O(0.1) v_{ms}$
(note that the units in the $v_{ms}$ panel are $10^8$~cm~s$^{-1}$
rather than the $10^7$~cm~s$^{-1}$ of the $v_{P\!f}$ and $v_z$ panels).
The slowness of these speeds demonstrates that even though much of the corona
is buoyantly unstable, the net accelerations are never more than a fraction
of gravity.

The relationship between $v_{P\!f}$ and $v_z$ changes sharply between the disk
core and the corona, as can be seen in panel f of Figure~\ref{fig:bspeed},
which shows their ratio.  In the disk core, they are entirely uncorrelated,
consistent with our conclusion that there are no genuine flows of either
in that part of the disk.   On the other hand, in the coronal region, $|z| > 2H$,
although not strictly proportional, more often than not they vary together,
with $v_{P\!f} \simeq (1$--$2)v_z$.  Where the undulatory Parker mode grows,
field lines develop upward bends and can carry gas upward with them.  However,
the curvature of the field lines permits the gas to slip diagonally downward
relative to the field, an effect explaining why the mean upward speed for
gas is smaller than that for field.

To sum up, we find that the undulatory Parker instability is the major driver of
the magnetic upwelling that has been so often noted in stratified shearing box
simulations. Most of the time, it is marginally stable in the core region
($|z|\lesssim 2H$), and unstable above it. Consistent upward mass and field
motions begin only at the lower boundary of the Parker instability, and both of
them are one order of magnitude slower than the magnetosonic speed. The upward
velocity of gas motion is generally half of the field flux velocity as the gas
may slide towards the field line valleys.

\subsection{Geometry of turbulent eddies}

The turbulence produced by the MRI is frequently described in the literature
as ``horizontal" because the initial work on shearing boxes
found that the kinetic energy of turbulent motions in the $x-$ and $y-$ directions,
corresponding to the radial and azimuthal directions if the box were placed in an
actual disk, was substantially greater than that in the $z-$, or vertical, direction.
For example, \citet{Hawley95} found that in their fiducial simulation of a
three-dimensional, but unstratified, shearing box (in this case, with initial magnetic
field uniform and vertical), $\langle \rho v_x^2 \rangle \simeq
\langle \rho (\delta v_y)^2 \rangle \simeq 4 \langle \rho v_z^2\rangle$, where
$\delta v_y$ is the y-velocity after subtracting out the shearing motion.
An initially azimuthal field also led to primarily horizontal motions, although
with somewhat larger amplitude in the radial than in the azimuthal direction.
Similarly, in a stratified shearing box that extended $\pm 2\sqrt{2}{\cal H}$
from the midplane, \citet{Stone96} found that the turbulence was predominantly
azimuthal: $\langle
\rho (\delta v_y)^2\rangle \simeq 6 \langle \rho v_x^2\rangle \simeq 20 \langle
\rho v_z^2\rangle$, with little dependence on whether the equation of state
was isothermal or adiabatic or on whether the initial field was vertical or
azimuthal.

However, the situation changes when stratified shearing boxes with
greater vertical extent are simulated, and the periodic boundary conditions
used by \citet{Stone96} are replaced by outflow boundary conditions.
\citet{Miller00} reported that in their simulations, in which the box extended
to $\pm 5\sqrt{2}{\cal H}$ from the midplane, the amplitude of vertical motion
near the midplane ($|z| \leq 2\sqrt{2}{\cal H}$) was almost as great as
the amplitude of the azimuthal turbulent motions:
$\langle \rho (\delta v_y)^2\rangle \simeq 1.2 \langle \rho v_z^2\rangle$,
nearly independent of whether the initial field was vertical or azimuthal.  Their
result for the ratio $\langle \rho v_z^2\rangle/\langle \rho v_x^2\rangle$
was more ambiguous: its value was $\simeq 0.5$ when the initial field was
azimuthal, but fell to $\simeq 0.27$ when it was vertical.
Our results are qualitatively similar to those of \citet{Miller00},
but indicate turbulent motions that are more nearly isotropic.
Averaging the standard resolution data over time and the region within $\pm 2H$
of the midplane, we have $\langle \rho (\delta v_y)^2\rangle \simeq
0.70\langle \rho v_x^2\rangle \simeq 1.2\langle \rho v_z^2\rangle$.  In the
highest-resolution simulation, the azimuthal motions grow slightly relative
to both the radial and vertical motions, but overall, all three velocity components
remain comparable in magnitude:
$\langle \rho (\delta v_y)^2\rangle \simeq
0.90\langle \rho v_x^2\rangle \simeq 1.6\langle \rho v_z^2\rangle$.
That such a qualitative contrast in the geometry of turbulent motions is
associated with the change from unstratified to stratified boxes strongly
suggests that its physical origin can be identified with the principal
physical contrast between the two cases: vertical gravity, which in this
context drives magnetic buoyancy.

\section{CONCLUSIONS \label{sec:conclusion}}
We have shown that in a stratified shearing box, unlike an unstratified one, the
magnetic saturation level does not diminish with increasing grid resolution.
Instead, at the resolutions used in our simulations, the Maxwell stress is
independent of the $x$ and $y$ resolution.  When
the vertical resolution increases, both the stress and the pressure grow.
Despite the nearly linear increase of both magnetic stress and total pressure
with resolution ($\langle B_r B_\phi\rangle,\langle P \rangle \propto (\Delta z)^{-1}$),
the ratio of stress to pressure is almost converged at this resolution
(as also found in the isothermal simulations of \citet{Davis09}),
increasing only slowly with finer vertical resolution.

The field structure of a stratified box is quite different from that of an
unstratified box.  At high altitude ($|z| > 2H$), the field is dominated
by its azimuthal component and forms smoothly distributed filamentary structures
with vertical thicknesses $\gtrsim H$.  In the core of the disk, the region
unstratified boxes are thought to mimic, the field is turbulent, but with
significantly larger scale features than found in unstratified simulations:
for simulations with identical resolution, the vertical correlation length of
$B_y$ is $\sim 3 $ times greater with vertical gravity than without.  In
addition, when gravity is present, the correlation length measured at this
resolution is much less dependent on gridscale, falling only
$\simeq 30\%$ when the resolution is doubled.

There are several ways to understand the contrast between magnetic field
behavior in the stratified and unstratified cases.  One is through
dimensional analysis.  Without vertical gravity, the ratio $c_s/\Omega$,
although well-defined, loses its physical significance.
Consequently, the only remaining significant lengthscale in the problem is
the gridscale.  As a result, the two lengthscales determined by the turbulence,
the field correlation length and $v_A/\Omega$, both track the gridscale and
approach zero as the resolution grows finer.  On the other hand, with vertical
gravity, $c_s/\Omega$ plays an explicit role in the turbulent dynamics, and
both the field correlation length and $v_A/\Omega$ can become associated
with it.  A related argument has been made by \citet{Vishniac09}.

A second approach ties the result more directly to gravitational dynamics.
As we have shown, buoyancy plays an important role in the character of
MRI-driven MHD turbulence in stratified shearing boxes.  Outside
the midplane region ($|z| \geq \pm (2$--$3)H$), $N^2_{mag}$ is consistently
negative, a signal that buoyant regions are continually being created.
Consequently, at any given time there is nearly always a rising magnetic
filament in that region.  Even inside the midplane region, vertical motions,
presumably excited by the buoyantly-driven pressure fluctuations on its edges,
are greater than they are in unstratified situations.  In notable contrast
to studies of both unstratified shearing boxes and stratified shearing
boxes with periodic vertical boundary conditions and containing only a few
vertical scale heights, we find that the turbulent motions are fully
three-dimensional, with the kinetic energy in vertical motion comparable
to that in either of the horizontal directions.

Because fully three-dimensional motion is essential to dynamo action (e.g.,
as reviewed by \cite{Cowling81} and first emphasized in this context by
\cite{Brand95}), the buoyant excitation of vertical motion is essential to
maintaining the magnetic field: as fluid elements rise and fall, they create
vertical field from horizontal.  The pressure scale height sets the characteristic
scale for vertical motions, linking the magnitude of the field (with
proportionality constant $\sim \sqrt{4\pi\rho}\Omega$) to this characteristic
lengthscale (see \cite{Vishniac09} for a proposed scaling relation).  At
saturation, field amplification by dynamo action is balanced by two field loss
mechanisms: dissipation and expulsion
of field from the box.  As previously shown by \citet{Hirose06}, the former
strongly dominates the latter when allowance is made for photon energy losses.

We have also used these simulations to identify the dynamical mechanism
responsible for the upward magnetic motions commonly seen in previous stratified
simulations: it is the undulatory Parker instability.  This mode is marginally
stable within the core region, but is almost always unstable outside that region.
True upward motions of both magnetic field and fluid begin at the instability
boundary, but are one order of magnitude slower than the magnetosonic speed.
Because the gas can slide along field lines and fall into field line valleys,
its mean upward velocity is generally only about half the Poynting velocity.


Although we have allayed fears that the true converged state of MRI-driven
MHD turbulence is zero stress, the question of numerical convergence of the
stratified case remains open.  We understand neither why the stress and
pressure are strongly dependent on vertical (but not horizontal) cell size
nor at what resolution this dependence may weaken.  The origin of the
large fluctuations as a function of time observed in box-integrated quantities
in these simulations is also unclear, although there are hints that their
magnitude may have to do with the radial width of the box \citep[e.g.,][]{FS09}.

Several physical questions also remain unanswered:
The pressure scale height certainly sets a physical lengthscale, but can we
understand more specifically how it determines the vertical correlation length
in the disk core's MHD turbulence?  It seems plausible that the field buoyancy
leads to larger scale features in the core region, as the field senses the
vertical outflow boundaries via the magnetic upwelling, but exactly how is
that communicated to the midplane? Lastly, one of the most striking features of
the disk corona is the quasi-regular alternation in sign of $B_y$: what causes
this alternation and what determines its characteristic lengthscale and timescale?

\acknowledgments

We thank Omer Blaes and Shane Davis for very useful discussions.
The authors acknowledge the Texas Advanced Computing Center (TACC) at The
University of Texas at Austin for providing HPC resources that have contributed
to the research results reported within this paper. Part of the work was also
performed on the Johns Hopkins Homewood High-Performance Computing Center cluster.
This work was partially supported by NASA ATP Grant NNG06GI68G and by
NSF Grant AST-0507455.


\clearpage

\clearpage


\begin{deluxetable}{c|ccccc}
\tabletypesize{\footnotesize}
\tablecolumns{6} \tablewidth{0pc}
\tablecaption{Properties of the simulations in the paper.}
\tablehead{\colhead{} & \colhead{STD32} &
\colhead{X64} & \colhead{Y128} & \colhead{Z512} & \colhead{DBLE}}

\startdata
resolution & $32\times64\times256$ & $64\times64\times256$  &
             $32\times128\times256$ & $32\times64\times512$ &
             $64\times128\times512$ \\

Run time (orbits) & 300 & 250 & 300 & 120 & 120 \\
$\Delta \Sigma_{max}/\Sigma~ (\%)$ & 2.5 & 2.0 & 2.0 & 1.1 & 1.4 \\
$\mathcal{H}=\langle\langle c_S/\Omega\rangle\rangle ~(H)$ & $1.65$ & $1.67$ & $1.67$ & $2.00$ & $1.94$ \\
$\langle\langle t_{cool} \rangle\rangle$ (orbits) & $8.0$ & $7.8$ &
$7.7$ & $6.6$ & $6.3$\\
$\langle\lambda_{z,~int}(B_y)\rangle_t ~(H)   $ & $0.31$ & $0.35$ & $0.34$ & $0.35$ & $0.34$ \\
$\langle\lambda_{z,~int}(B_y;y=0)\rangle_t~(H)$ & $0.31$ & $0.36$ & $0.33$ & $0.35$ & $0.33$ \\
$\langle\lambda_{z,~FWHM}(B_y)\rangle_t~(H)$ & $0.26$ & $0.25$ & $0.25$ &
$0.23$ & $0.22$ \\
\hline
$\langle\langle-B_xB_y/4\pi\rangle\rangle/P_0$        & $0.029$ & $0.030$ &
                                                        $0.031$ & $0.064$ & $0.060$ \\

$\langle\langle\rho v_x \delta v_y\rangle\rangle/P_0$ & $6.7\times10^{-3}$ & $6.7\times10^{-3}$ &
                                                        $7.0\times10^{-3}$ & $0.015$ & $0.013$ \\

$\langle\langle B^2/8\pi\rangle\rangle/P_0$   & $0.11$ & $0.11$ &
                                                $0.11$ & $0.24$ & $0.20$ \\
$\langle\langle B^2_x/8\pi\rangle\rangle/P_0$ & $9.1\times10^{-3}$ & $9.3\times10^{-3}$ &
                                                $0.010$ & $0.025$  & $0.024$ \\
$\langle\langle B^2_y/8\pi\rangle\rangle/P_0$ & $0.095$ & $0.090$ &
                                                $0.096$ & $0.21$  & $0.16$ \\
$\langle\langle B^2_z/8\pi\rangle\rangle/P_0$ & $5.9\times10^{-3}$ & $6.8\times10^{-3}$ &
                                                $6.5\times10^{-3}$ & $0.014$ &
                        $0.015$ \\

$\langle\langle\rho\delta v^2/2\rangle\rangle/P_0$  & $0.029$ & $0.031$ &
                                                      $0.031$ & $0.061$ & $0.057$ \\
$\langle\langle\rho v^2_x/2\rangle\rangle/P_0$      & $0.012$ & $0.012$ &
                                                      $0.012$ & $0.024$ & $0.021$ \\
$\langle\langle\rho\delta v^2_y/2\rangle\rangle/P_0$& $0.010$ & $0.011$ &
                                                      $0.011$ & $0.023$ & $0.023$ \\
$\langle\langle\rho v^2_z/2\rangle\rangle/P_0$ &      $7.4\times10^{-3}$ & $8.0\times10^{-3}$ &
                                                      $7.6\times10^{-3}$ &
                              $0.014$ & $0.014$ \\

\hline
$\langle\langle-B_xB_y/4\pi\rangle/P(0)\rangle$ & $3.9\times10^{-3}$ &
                                                  $4.0\times10^{-3}$ &
                                                  $4.1\times10^{-3}$ &
                          $7.1\times10^{-3}$ &
                                                  $7.3\times10^{-3}$ \\

$\langle\langle\rho v_x \delta v_y\rangle/P(0)\rangle$ & $8.8\times10^{-4}$ &
                                                         $8.8\times10^{-4}$ &
                                                         $9.2\times10^{-4}$ &
                             $1.7\times10^{-3}$ &
                                                         $1.5\times10^{-3}$ \\

$\langle\langle B^2/8\pi\rangle/P(0)\rangle$   & $0.014$ & $0.014$ &
                                                 $0.015$ & $0.040$ &
                                                 $0.024$ \\
$\langle\langle B^2_x/8\pi\rangle/P(0)\rangle$ & $1.2\times10^{-3}$ &
                                                 $1.2\times10^{-3}$ &
                                                 $1.3\times10^{-3}$ &
                         $2.7\times10^{-3}$ &
                                                 $2.9\times10^{-3}$ \\
$\langle\langle B^2_y/8\pi\rangle/P(0)\rangle$ & $0.012$ &
                                                 $0.012$ &
                                                 $0.013$ &
                         $0.036$ & $0.020$ \\
$\langle\langle B^2_z/8\pi\rangle/P(0)\rangle$ & $7.7\times10^{-4}$ &
                                                 $8.8\times10^{-4}$ &
                                                 $8.5\times10^{-4}$ &
                         $1.4\times10^{-3}$ &
                                                 $1.9\times10^{-3}$ \\

$\langle\langle\rho\delta v^2/2\rangle/P(0)\rangle$  & $3.8\times10^{-3}$ &
                                                       $4.1\times10^{-3}$ &
                                                       $4.0\times10^{-3}$ &
                               $6.7\times10^{-3}$ &
                                                       $6.9\times10^{-3}$ \\
$\langle\langle\rho v^2_x/2\rangle/P(0)\rangle$ &      $1.6\times10^{-3}$ &
                                                       $1.6\times10^{-3}$ &
                                                       $1.6\times10^{-3}$ &
                               $2.7\times10^{-3}$ &
                                                       $2.6\times10^{-3}$ \\
$\langle\langle\rho\delta v^2_y/2\rangle/P(0)\rangle$& $1.3\times10^{-3}$ &
                                                       $1.4\times10^{-3}$ &
                                                       $1.4\times10^{-3}$ &
                               $2.5\times10^{-3}$ &
                                                       $2.8\times10^{-3}$ \\
$\langle\langle\rho v^2_z/2\rangle/P(0)\rangle$ &      $9.7\times10^{-4}$ &
                                                       $1.0\times10^{-3}$ &
                                                       $1.0\times10^{-4}$ &
                               $1.5\times10^{-4}$ &
                                                       $1.6\times10^{-4}$ \\
\hline
$\langle\langle-B_xB_y/4\pi\rangle/\langle P \rangle\rangle$ & $0.020$ & $0.021$ &
                                                               $0.021$ & $0.029$ & $0.029$ \\

$\langle\langle\rho v_x \delta v_y\rangle/\langle P\rangle\rangle$ & $4.6\times10^{-3}$ & $4.6\times10^{-3}$ &
                                                                     $4.7\times10^{-3}$
                                     &
                                     $6.8\times10^{-3}$
                                     & $6.3\times10^{-3}$ \\

$\langle\langle B^2/8\pi\rangle/\langle P\rangle\rangle$   & $0.075$ & $0.073$ &
                                                             $0.076$ & $0.13$  &
                                 $0.098$ \\
$\langle\langle B^2_x/8\pi\rangle/\langle P\rangle\rangle$ & $6.1\times10^{-3}$ & $6.4\times10^{-3}$ &
                                                             $6.8\times10^{-3}$
                                 & $0.011$ & $0.012$ \\
$\langle\langle B^2_y/8\pi\rangle/\langle P\rangle\rangle$ & $0.064$ & $0.062$ &
                                                             $0.065$ & $0.11$ &
                                 $0.079$ \\
$\langle\langle B^2_z/8\pi\rangle/\langle P\rangle\rangle$ & $4.0\times10^{-3}$ & $4.6\times10^{-3}$ &
                                                             $4.3\times10^{-3}$
                                 &
                                 $5.9\times10^{-3}$
                                 & $7.5\times10^{-3}$ \\

$\langle\langle\rho\delta v^2/2\rangle/\langle P\rangle\rangle$  & $0.020$ & $0.021$ &
                                                                   $0.021$ & $0.028$ & $0.028$ \\
$\langle\langle\rho v^2_x/2\rangle/\langle P\rangle\rangle$ &      $8.2\times10^{-3}$ & $8.3\times10^{-3}$ &
                                                                   $8.5\times10^{-3}$ & $0.011$ & $0.010$ \\
$\langle\langle\rho\delta v^2_y/2\rangle/\langle P\rangle\rangle$ &$6.8\times10^{-3}$ & $7.6\times10^{-3}$ &
                                                                   $7.3\times10^{-3}$ & $0.010$ & $0.011$ \\
$\langle\langle\rho v^2_z/2\rangle/\langle P\rangle\rangle$ &      $5.1\times10^{-3}$ & $5.5\times10^{-3}$ &
                                                                   $5.2\times10^{-3}$
                                   &
                                   $6.3\times10^{-3}$
                                   & $6.7\times10^{-3}$ \\

\hline
$\langle\langle-B_xB_y/4\pi\rangle/\langle B^2/8\pi\rangle\rangle$ & $0.27$ &
                                                                     $0.28$ &
                                                                     $0.28$ &
                                     $0.21$ &
                                     $0.31$\\
$\langle\langle-B_xB_y/4\pi\rangle/\langle\rho v_x\delta v_y\rangle\rangle$ &
                                                                            $4.6$ &
                                                                            $4.7$ &
                                        $4.5$ &
                                        $4.4$ &
                                        $4.8$\\
$\langle\langle-B_xB_y/4\pi\rangle\rangle/\langle\langle B^2/8\pi\rangle\rangle$
& $0.26$ & $0.27$ & $0.28$ & $0.27$ & $0.30$\\
$\langle\langle-B_xB_y/4\pi\rangle\rangle/\langle\langle\rho v_x\delta
v_y\rangle\rangle$ & $4.3$ & $4.5$ & $4.4$ & $4.3$ & $4.6$ \\
\enddata \label{tab:1}
\end{deluxetable}

\end{document}